# Effectiveness of the product stabilization in direct three-body recombination

Ekaterina V. Ermolova, Lev Yu. Rusin*

* Corresponding author. Tel.: +07 499 1374104, fax: +07 499 1378258.
*E-mail address:* rusin@chph.ras.ru (L.Yu. Rusin).

*V.L. Talroze Institute of Energy Problems of Chemical Physics, The Russia Academy of Sciences, Leninskii prospect 38, Bldg. 2, Moscow 119334, Russia*

Abstract

The optimal kinematic parameters for the formation of the most stabilized products in direct three-body recombination $Cs^+ + Br^- + R \rightarrow CsBr + R$ with R = Hg, Xe, Kr are determined by the deformed polyhedron method for the third body energy and the ion encounter energy ranging from 1 to 10 eV. The stabilization effectivity functions are obtained for each atom, and the mean effectivities of the atoms in the formation of products with the lowest internal energy are calculated. The Xe and Hg atoms are almost equally effective in energy removal from the recombining pair due to high hardness of the repulsion potentials in the interactions of these atoms with the ions. The Kr atom effectivity is much less because of much softer repulsion in the corresponding pairwise interactions. The dynamical mechanism of the third body collisions with the ions affects greatly the energy transfer effectiveness.

## 1. Introduction

The elementary reactions of direct three-body recombination play an important role in many processes observed in various media and complex chemical systems such as low temperature plasma, combustion of gas mixtures, and so on. For instance, the analysis of Ref. [1] has revealed that about half of the 196 principal elementary reactions in combustion processes [2] are recombination reactions and reverse reactions of collision-induced dissociation of molecules. Direct three-body recombination is one of the main types of elementary reactions responsible for the concentrations of atoms, atomic ions, and small radicals in many chemical processes. However, the mechanisms of such elementary reactions have been explored rather poorly. The corresponding experimental data are mainly the rate constants, and they are interpreted, as a rule, from the viewpoint of a simultaneous collision of all the three particles leading to the formation of the final product. The ample experimental information collected on the rate constants of termolecular recombination reactions [3,4] does not include data on the process dynamics or on energy redistribution mechanisms in three-body collisions. The reason is that the rate constant of such a reaction is the result of multiple averaging of the process probability over the distributions of the kinematic parameters of the three-body collision yielding recombination. Due to the presence of many parameters involved in the averaging, one has to study the three-body process dynamics and various quantities describing this dynamics to comprehend the mechanism of the interaction of the third body with the recombining particles. To be more precise, one has to examine the quantities that characterize the energy transfer process in much more detail than the rate constant of the reaction does. Such quantities are exemplified by the energy states of the product molecules.

The essential but almost unexplored aspects of the recombination mechanism include the dynamics of stabilization of the nascent molecule by the third body and the effectiveness of various particles as stabilizers of the product molecule. What is used in chemical kinetics as a parameter for comparing effectivities of the third bodies in stabilizing recombination products is the termolecular rate constant. For instance, Ref. [5] presents the rate constants of iodine atom recombination measured at 20° C, the third bodies (quite different in their masses) being the five rare gas atoms. These rate constants increase from $1.73 \times 10^{-32}$ ml.$^2$ molecules$^{-2}$ s$^{-1}$ for helium to $3.44 \times 10^{-32}$ ml.$^2$ molecules$^{-2}$ s$^{-1}$ for xenon, which could indicate the role of the mass of the rare gas atom. Ref. [6] gives somewhat smaller values for the same rate constants which however also increase from $0.67 \times 10^{-32}$ ml.$^2$ molecules$^{-2}$ s$^{-1}$ for He to $2.99 \times 10^{-32}$ ml.$^2$ molecules$^{-2}$ s$^{-1}$ for Xe. One should note here that the polarizabilities of the rare gas atoms also grow from 0.205 Å$^3$ for He to 4.01 Å$^3$ for Xe [7]. If there does take place the two-stage recombination mechanism including complex formation I + R → IR and the subsequent reaction IR + I → $I_2$ + R, the increase in the rate constant in the rare gas sequence could be due to an increment in the polarizabilities, at least partially.

The simple model of an encounter of two hard spheres implies that efficiency of the kinetic energy exchange is determined, to a great extent, by the mass ratio of the colliding particles. However, such a straightforward model cannot be applied to three-body reaction where two recombining ions or atoms form a molecule to be stabilized by the third body. The reason is that stabilization of the recombination product can occur via the interaction of the third body with any of the recombining particles, with two particles in any sequence, or with both the particles simultaneously. This compels one to consider the dynamics of the whole three-body process. In the papers [8–17], general dynamics as well as some aspects of detailed dynamics of three-body recombination of the Cs$^+$ and Br$^-$ ions in the presence of the Kr, Xe, or Hg atoms were examined (recall that in contrast with general dynamics, also known as *statistical dynamics*, researches of *detailed dynamics* are characterized by no averaging or minimal averaging over kinematic parameters). These studies proved that in any adequate recombination model, one has to take into account the interaction potentials of all the three particles, i.e., the potential energy surface (PES) that governs the three-body collision. The PES structure depends on the interaction parameters of all the three partners.

Thus, the dynamical mechanism of the elementary process of recombination and the effectiveness of the third body as a stabilizer of the nascent molecule should depend on the PES structure as well as on the collection of the kinematic parameters of the three-body collision. These parameters determine a trajectory of the representative point on the PES (if one uses the quasiclassical trajectory method) and, consequently, the outcome of the interaction of the recombining pair with the third body. In general, the masses of the partners are among the kinematic parameters. Compared with two-body collisions, three-body interactions are characterized by a more complicated collection of the kinematic parameters [10–17].

Unfortunately, the current experimental techniques do not enable one to determine the values of the kinematic parameters yielding the formation of a molecule in a prescribed energy state or with the minimal internal energy in three-body recombination. However, if an adequate PES of the reverse reaction of collision-induced dissociation of the molecule is known, one can explore numerically the interaction of the nascent molecule with the third body and figure out the main factors responsible for stabilization of the recombination products with a sufficient accuracy. Extensive trajectory simulations of collisions of the CsBr molecule with R = Kr, Xe, Hg were carried out on certain potential energy surfaces (note that this molecule has an ionic bond and dissociates into the Cs$^+$ and Br$^-$ ions), see e.g. Refs. [10,18–22]. These simulations showed the possibility of reproducing quantitatively correctly the manifold experimental data obtained in crossed molecular beams [10,18–21]. The Kr, Xe, and Hg atoms have therefore been selected as the third bodies to be compared in the *reverse* processes of direct three-body recombination of the $Cs^+$ and $Br^-$ ions. The masses of these atoms in the pairs (Hg, Xe) and (Xe, Kr) differ by a factor of about 1.5, and the PESs of the CsBr + R systems differ in their structures. These facts are expected to allow one to determine the effects of the third body mass and the PES structure on the effectiveness of recombination product stabilization.

The present paper studies effectivities of the atoms R = Hg, Xe, Kr as the third bodies in stabilization of the products of three-body recombination of the $Cs^+$ and $Br^-$ ions undergoing central encounters:

$$Cs^+ + Br^- + R \rightarrow CsBr + R, \quad R = Hg, Xe, Kr. \qquad (1)$$

The goal of this research is to figure out how the masses of the third bodies and the structures of the corresponding PESs affect the dynamics of recombination product stabilization. The article is confined with *central* (head-on) encounters of the ions, i.e., encounters with zero impact parameter. Within this framework, stabilization effectiveness of the product molecule is determined by energy transfer only, and the process is not complicated by angular momentum transfer. The peculiarities of the effectivities of the third bodies for non-central encounters of the ions are to be examined separately (cf. Refs. [12,17]). Moreover, all the three particles are assumed to arrive at the zone of strong repulsive interaction simultaneously (*one-stage* recombination). The effectivities of the third bodies in two-stage recombination processes are also beyond the present study (cf. Ref. [17]).

The paper is organized as follows. After this introduction, in Section 2, the problem of optimizing the kinematic parameters in the case of central encounters of the ions is discussed. The optimal kinematic parameters are expected to ensure the deepest stabilization of the product molecules. The same section describes the potential energy surfaces used for each of the third bodies Hg, Xe, Kr. The minimal internal energies of CsBr computed for the ion encounter energy from 1 to 10 eV and the third body energy from 1 to 10 eV as well as the corresponding optimal impact parameters of the third body are presented in Section 3. The optimal orientation angles of the initial velocities of the particles are also considered in Section 3. The results obtained are discussed in Section 4 where the third bodies in question are compared in their effectivity in stabilizing the recombination products and the most probable explanation of the differences found is given. The main conclusions of the work are contained in Section 5.

## 2. Kinematic parameter optimization and the potential energy surfaces

Within the quasiclassical trajectory method, the outcome of a collision of the reagents on a given PES governing their interaction is completely determined by the collection of kinematic parameters. In the setup under consideration, where one-stage direct three-body recombination of the two ions $Cs^+$ and $Br^-$ is examined in the simplest case of central encounters of the ions, the kinematic parameters are the ion encounter energy $E_i$, the third body energy $E_R$ (the relative kinetic energy of the recombining pair and the third body R), the impact parameter $b_R$ of the third body with respect to the center-of-mass of the ions, the polar angles $\Theta$ and $\Phi$ of the initial relative velocity of the ions with respect to the initial relative velocity of the ion pair and the third body, and the masses of the three particles involved. A more detailed description of these kinematic parameters is presented in the papers [10–17]. The values $b_R = 0$, $\Theta = 0$ correspond to the collinear configuration $Cs^+ - Br^- - R$.

Given a potential energy surface, the internal energy of the product molecule CsBr is a function $E_{int}$ of the eight kinematic parameters just listed. The definition domain of this function is the region of kinematic parameter values that lead to recombination. The problem of determining the kinematic parameter values yielding the deepest stabilization of the recombination product is tantamount to finding the global minimum of the function $E_{int}$. In the present paper, the energies $E_i$ and $E_R$ are always assumed to be fixed in the minimization procedure, the masses of the recombining ions are assumed to be equal to the actual isotope-averaged masses of the Cs and Br atoms, and the mass of the third body is equal to the actual isotope-averaged mass of one of the atoms Hg, Xe, or Kr. In this setting, the deepest stabilization of the product molecule by the third body corresponds to the global minimum of the function $E_{int}$ of the three variables $b_R$, $\Theta$, and $\Phi$.

By means of integrating the equations of motion for any collection of the values of $b_R$, $\Theta$, and $\Phi$ (for the values of the other kinematic parameters fixed beforehand), one can either compute the corresponding internal energy $E_{int}$ of the CsBr molecule or reveal that the collection in question does not lead to recombination. It is clear that the function $E_{int} = E_{int}(b_R, \Theta, \Phi)$ cannot be represented by any "reasonably short" analytic expression, and it is not an easy task to calculate the partial derivatives of this function. It is therefore hardly possible to find the minimum of this function by well-known gradient methods. On the other hand, the internal energy of the product molecule can be minimized with respect to the kinematic parameters by *direct search methods*, or methods of zero order [23]. There exist many various gradient-free methods for searching the global extrema of functions of several variables [23–27]. To the best of the authors' knowledge, such methods have not been applied yet (on any noticeable scale) to the problems of dynamics of elementary processes.

In Ref. [28], an analysis of applicability of several optimization algorithms to studies of direct three-body recombination of ions has been carried out. This analysis has revealed that the optimization problems encountered in such studies are best suited by the so-called *simplex methods* which are based on a movement of a simplex in the space of the arguments (recall that a simplex in an *n*-dimensional space is a polytope with $n+1$ vertices) via reflections of a vertex of the simplex through the center-of-mass of the opposite face. The first simplex algorithm was proposed by Spendley, Hext, and Himsworth [29]. Its advanced modification where deformed polyhedra are considered and so-called expansion and contraction moves are used along with reflections is probably the most accurate (and at the same time reasonably fast) procedure for minimizing the internal energy of the recombination products [28]. This modification was invented by Nelder and Mead [30], its modern careful and complete description is presented in Ref. [31] (see also Refs. [11,15,16,23–28] for various accounts). Like the previous articles [11,15,16], the present paper employs the Nelder–Mead method (to be also called the *deformed polyhedron method*) to minimize the function $E_{int}(b_R, \Theta, \Phi)$. The technique of detecting and eliminating the so-called false (local but not global) minima in this optimization problem is proposed and discussed in Ref. [15]. The search is performed within the domain given by the inequalities $0 \le b_R \le 3$ au (or $0 \le b_R \le 3.5$ au), $0 \le \Theta \le \pi$, $0 \le \Phi < 2\pi$ (test searches have proven that $E_{int}$ cannot attain the minimum at $b_R > 3.5$ au). The argument values corresponding to the minimum will be said to be *optimal* and denoted by $b_R^*$, $\Theta^*$, $\Phi^*$.

As was already pointed out above, the recombination process itself is simulated in this study via quasiclassical trajectories on three different PESs whose adequacy is confirmed by reproducing the experimental data (obtained in crossed molecular beams) on the interaction in the systems CsBr + Kr [10], CsBr + Xe [18–20], and CsBr + Hg [21]. The details of the trajectory simulation procedure (in particular, the choice of initial conditions) are described in the preceding papers [8,10,11,13,14,16] devoted to one-stage direct three-body recombination of ions in central encounters. The potential energy surfaces used in this article and in the previous studies of the reactions (1) [8–17,28] deserve some discussion.

For all the three systems $Cs^+ + Br^- + R$, the PESs in question are constructed on the basis of the pairwise interaction potentials $U(r)$ of the particles, the interionic potential being modeled by the truncated Rittner expression [32,33].

For R = Kr and Xe, each of the potentials R–$Cs^+$ and R–$Br^-$ is assumed to be the sum of the repulsive Born–Mayer wall, the "charge – induced dipole" interaction, and the van der Waals interaction in the London approximation. The potential energy surface for R = Kr or Xe is given by the following analytic expression in atomic units:

$$V(r_1, r_2, r_3) = A_2 \exp\left(-\frac{r_2}{\rho_2}\right) - \frac{1}{r_2} - \frac{\alpha_2 + \alpha_3}{2r_2^4} - \frac{C_2}{r_2^6}$$
$$+ A_1 \exp\left(-\frac{r_1}{\rho_1}\right) + A_3 \exp\left(-\frac{r_3}{\rho_3}\right) - \frac{C_1}{r_1^6} - \frac{C_3}{r_3^6} \quad (2)$$
$$- \frac{\alpha_1}{2}\left(\frac{1}{r_1^4} + \frac{1}{r_3^4} + \frac{r_2^2 - r_1^2 - r_3^2}{r_1^3 r_3^3}\right)\left(1 - \frac{\alpha_2 + \alpha_3}{r_2^3}\right)^2.$$

Here $r_1$, $r_2$, and $r_3$ are the internuclear distances in the pairs $Cs^+$–R, $Cs^+$–$Br^-$, and $Br^-$–R, respectively, $A_i$ and $\rho_i$ are the Born–Mayer parameters of the corresponding diatomic fragments, $C_i$ denote the dispersion constants ($1 \leq i \leq 3$), while $\alpha_1$, $\alpha_2$, $\alpha_3$ are the polarizabilities of R, $Cs^+$, $Br^-$, respectively. The cross term in Eq. (2) describes the polarization interaction in the system of the R atom and the $Cs^+$–$Br^-$ dipole.

For R = Hg, each of the potentials Hg–$Cs^+$ and Hg–$Br^-$ is assumed to be the sum of the Lennard-Jones (12, 6) potential and the "charge – induced dipole" interaction, the LJ parameters corresponding to the system of the mercury atom and the rare gas atom isoelectronic to $Cs^+$ or $Br^-$, i.e., Xe or Kr. The analytic expression of the potential energy surface for R = Hg in atomic units is

$$V(r_1, r_2, r_3) = A_2 \exp\left(-\frac{r_2}{\rho_2}\right) - \frac{1}{r_2} - \frac{\alpha_2 + \alpha_3}{2r_2^4} - \frac{C_2}{r_2^6}$$
$$+ D_1\left[\left(\frac{r_{01}}{r_1}\right)^{12} - 2\left(\frac{r_{01}}{r_1}\right)^6\right] - \frac{\alpha_1}{2r_1^4} \quad (3)$$
$$+ D_3\left[\left(\frac{r_{03}}{r_3}\right)^{12} - 2\left(\frac{r_{03}}{r_3}\right)^6\right] - \frac{\alpha_1}{2r_3^4}.$$

Here the quantities $r_1$, $r_2$, $r_3$, $A_2$, $\rho_2$, $C_2$, $\alpha_1$, $\alpha_2$, $\alpha_3$ have the same meaning as in Eq. (2), $D_1$ and $D_3$ are the well depths of the LJ potentials while $r_{01}$ and $r_{03}$ are the equilibrium distances of these potentials.

The parameters $A_i$, $\rho_i$, $C_i$, $D_i$, $r_{0i}$ in Eqs. (2) and (3) characterize the interaction potentials $U(r)$ of the diatomic fragments (one restores the diatomic potentials from Eqs. (2) and (3) by setting two of the distances $r_1$, $r_2$, $r_3$ to be sufficiently large). In the literature, there are presented many collections of values for these parameters, the values of some parameters given in different works differ drastically (by up to 50%). The values derived from the data of dynamical experiments (concerning e.g. ion mobilities or scattering in crossed molecular beams) are the most reliable ones. The values used in Refs. [8–17,28] and in the present paper are based on the data of Refs. [32–37] and compiled in Table 1. For the polarizabilities of the particles, the values $\alpha_{Cs^+} = 16.48$ [32], $\alpha_{Br^-} = 32.46$ [32], $\alpha_{Kr} = 16.8$ [38], $\alpha_{Xe} = 27.2$ [38], $\alpha_{Hg} = 34$ au$^3$ [7] are adopted.

**Table 1**
The potential parameters of the systems Cs$^+$ + Br$^-$ + R (R = Kr, Xe, Hg) in atomic units.

| The parameter | R = Kr | R = Xe | R = Hg | The references |
|---|---|---|---|---|
| $A_2$ | | 127.5 | | [32,33] |
| $\rho_2$ | | 0.7073 | | |
| $C_2$ | | 87.36 | | |
| $A_1$ | 796 | 318.5 | | [34] |
| $\rho_1$ | 0.5281 | 0.6494 | | |
| $C_1$ | 247.1 | 490 | | |
| $A_3$ | 62.3 | 62.84 | | [35] |
| $\rho_3$ | 0.723 | 0.877 | | |
| $C_3$ | 317 | 297.3 | | |
| $D_1$ | | | 0.0011 | [36] |
| $r_{01}$ | | | 7.75 | |
| $D_3$ | | | 0.00081 | [37] |
| $r_{03}$ | | | 7.56 | |

The PES for the Cs$^+$ + Br$^-$ + Hg system used in Ref. [21] contained a cross term similar to the cross term in Eq. (2). In Refs. [8–11,15,28] and in the present paper, the PES of Eq. (3) without a cross term is used for R = Hg. Test calculations have shown that the cross term of Ref. [21] does not affect the recombination dynamics almost at all.

For each of the third bodies R = Hg, Xe, Kr and each pair $(E_i, E_R)$ of the values of the ion encounter energy $E_i$ and the third body energy $E_R$, denote by $E_m(E_i, E_R)$ the minimum of the product internal energy $E_{int} = E_{int}(b_R, \Theta, \Phi)$. In other words, $E_m$ is the lowest internal energy (found by the Nelder–Mead method [30]) of the product CsBr of one-stage direct three-body recombination of Eq. (1) for given values of $E_i$ and $E_R$ under the assumption that encounters of the ions are central. Each of the energies $E_i$ and $E_R$ in the present study has been sampled in the interval between 1 and 10 eV with a step of 1 eV. The two-dimensional array $E_m = E_m(E_i, E_R)$ thus obtained characterizes the effectiveness of the third body R in stabilization of the nascent CsBr molecule. One may call this array the *effectivity function* of the third body R. As a single number characterizing the effectiveness of the third body one can use the *mean effectivity*

$$S_R = \frac{1}{N} \sum_{E_i, E_R} E_m(E_i, E_R), \qquad (4)$$

where $N = 100$ is the number of the pairs $(E_i, E_R)$ sampled. The smaller $S_R$, the *higher* is the effectiveness of the third body R.

To "decouple" the effects of the potential energy surface from those of the mass of the third body, on each of the three PESs for the Cs$^+$ + Br$^-$ + R systems considered, R = Hg, Xe, Kr, trajectory simulation of direct three-body recombination has been carried out not only for the "natural" case where the mass of the third body is equal to that of the R atom but also for the third bodies with masses equal to the masses of the two other atoms (see Section 4 below for details). In the situation where the PES corresponds to the system Cs$^+$ + Br$^-$ + R but the third body has the mass of the P atom, P ≠ R, one can also compute the effectivity function and the mean effectivity similarly to Eq. (4). This mean "cross-effectivity" will be denoted in the sequel by $^PS_R$.

## 3. Calculation results

The arrays $E_m(E_i, E_R)$ of the minimal internal energy $E_m$ of the CsBr molecules stabilized by a collision with the third body R = Hg, Xe, Kr are shown in Figs. 1–3, both the ion encounter energy $E_i$ and the third body energy $E_R$ ranging between 1 and 10 eV. To each point $(E_i, E_R)$, there correspond the optimal kinematic parameters $b_R^*$, $\Theta^*$, $\Phi^*$ yielding the deepest stabilization of the recombination product. As one can see in Figs. 1 and 2, the energies $E_m$ for the heavy atoms Hg and Xe never exceed 0.7 eV. For R = Kr, the minimal internal energies $E_m$ of the stabilized molecules lie in the interval from 0 to 2.6–2.7 eV (Fig. 3). This implies that stabilization by Kr atoms is much less effective than that by Xe and Hg atoms. Of the three effectivity functions $E_m(E_i, E_R)$ presented in Figs. 1–3, the function for R = Hg (Fig. 1) exhibits the simplest structure: the energy $E_m$ decreases gradually as the energy $E_R$ of the collision of the mercury atom with the ion pair grows from 1 to 10 eV. For $E_R$ in a wide interval between 4–5 and 10 eV, the minimal internal energy of CsBr does not exceed 0.1 eV. For the largest values of $E_R$ sampled, a slight rise in $E_m$ (still no greater than 0.1 eV) occurs. Another peculiarity of the effectivity function for R = Hg is that $E_m$ does not depend on the ion encounter energy $E_i$ almost at all. As $E_R$ decreases from 4–5 to 1 eV, product stabilization in collisions with Hg atoms gets less deep, and the minimal internal energy of CsBr increases to its largest value of ≈0.7 eV.

Fig. 2 shows that the shape of the effectivity function $E_m(E_i, E_R)$ for R = Xe is similar to that for R = Hg although its behavior for R = Xe is somewhat more complicated. In the energy range $3 \leq E_R \leq 7$ eV, the value of $E_m$ does not exceed 0.1 eV, but $E_m$ rises to ≈0.7 eV as the third body energy $E_R$ decreases to 1 eV. In contrast with the effectivity function in Fig. 1, an increase in $E_R$ above 7 eV for R = Xe leads to an enhancement in the minimal internal energy of CsBr to ≈0.5 eV. In the region $7 \leq E_R \leq 10$ eV, the energy $E_m$ turns out to depend strongly on the ion encounter energy $E_i$. As $E_i$ grows from 7 to 10 eV, the minimal internal energy of the product molecules becomes smaller than 0.1 eV although it does not reach nearly zero values typical for $E_R$ close to 5 eV. The mean effectivities $S_{Hg}$ and $S_{Xe}$ computed according to Eq. (4) are equal to 0.16 and 0.15 eV, respectively. The Xe and Hg atoms as energy acceptors are therefore of comparable effectiveness.

The effectivity function $E_m(E_i, E_R)$ for R = Kr shown in Fig. 3 is much more intricate than those for R = Hg and Xe. This function differs drastically from those presented in Figs. 1 and 2 not only by a much wider range of the $E_m$ values but also in the structure of the dependence of $E_m$ on $E_i$ and $E_R$. For low energies $E_R \leq 5$ eV, the minimal internal energy of the CsBr molecules does not exceed 0.4 eV and does not depend on the energy $E_i$ almost at all. As $E_R$ decreases from 3 to 1 eV, a slight increase in $E_m$ is observed. The region on the $(E_i, E_R)$ plane where an extremely deep stabilization of the product molecules is achieved ($E_m \leq 0.1$ eV) is much narrower than that for R = Hg and Xe. In contrast with the effectivity functions $E_m(E_i, E_R)$ for R = Hg and Xe considered above, an increase in $E_R$ from 4–4.5 eV to 8–9 eV yields more weakly stabilized molecules and an almost monotonous rise in the $E_m$ values. Moreover, for high energies $E_R \geq 9$ eV, the behavior of the effectivity function for R = Kr is significantly different compared with the graphs shown in Figs. 1 and 2. Here $E_m$ exhibits a strong dependence on the ion encounter energy $E_i$, especially for high $E_i$ values. At $E_R = 10$ eV and $9 \leq E_i \leq 10$ eV, the energy $E_m$ attains its maximum of 2.6–2.7 eV. The mean effectivity $S_{Kr}$ for R = Kr is equal to 0.71 eV which is much larger than the mean effectivities of the Hg and Xe atoms.

One of the important kinematic parameters determining the particle collision dynamics in direct three-body recombination is the impact parameter $b_R$ of the third body with respect to the center-of-mass of the recombining pair. Figs. 4–6 show the dependence of the optimal impact parameter $b_R^*$ corresponding to the deepest stabilization of the CsBr molecules on the energies $E_i$ and $E_R$ for R = Hg, Xe, Kr. For the heaviest third body R = Hg, over a sizeable region on the $(E_i, E_R)$ plane, the minimal internal energy of the molecule is attained at impact parameters $b_R^*$ lying between 1.8 and 3 au (Fig. 4). Large optimal impact parameters $b_R^* \geq 2.5$ au occur for not so high energies $E_R$ of the third body. Such values of $E_R$ bound the distances of the closest approach of the third body to the ions and consequently the repulsion energy between the third body and the ions. As $E_R$ increases from 1 to 10 eV, the impact parameter $b_R^*$ first decreases slightly and then (as $E_R$ passes ≈6 eV) drops abruptly to values smaller than 0.5 au (for low energies $E_i$ of ion encounters). The dependence of $b_R^*$ on $E_i$ is rather strong for high energies $E_R$.

For R = Xe (Fig. 5), the topography of the function $b_R^*(E_i, E_R)$ is similar, but the decrease in $b_R^*$ as $E_R$ grows is steeper than for R = Hg. However, the region on the $(E_i, E_R)$ plane where small impact parameters $b_R$ are required for deep stabilization of the CsBr molecules is much wider for R = Xe than for R = Hg.

The structure of the dependence of $b_R^*$ on $E_i$ and $E_R$ for R = Kr (Fig. 6) is quite different. For the same ranges of the two collision energies, the impact parameters that correspond to the deepest stabilization of the recombination product are significantly smaller on the whole. As $E_R$ increases, the optimal impact parameter $b_R^*$ decreases much faster (from $b_R^* \approx 1.8$ au at $E_R = 1$ eV to $b_R^* \leq 0.5$ au at $E_R = 3$ eV) than for the heavier third bodies R = Xe and Hg. The region of very small impact parameters $b_R^*$ (less than 0.5 au) stretches from $E_R = 3$ eV to $E_R = 10$ eV (the inequality $b_R^* < 0.5$ au is almost independent of the ion encounter energy $E_i$).

The angles $\Theta$ and $\Phi$ describing the mutual orientation of the initial velocities of the collision partners also affect the formation of CsBr products with minimal internal energies. For R = Hg and Xe, there are no noticeable regular peculiarities in the functions $\Theta^*(E_i, E_R)$ and $\Phi^*(E_i, E_R)$. For R = Hg, the optimal angles $\Theta^*$ and $\Phi^*$ range in the intervals $50° \leq \Theta^* \leq 100°$ and $0° \leq \Phi^* \leq 150°$. For R = Xe, these angles range in the intervals $50° \leq \Theta^* \leq 100°$ and $100° \leq \Phi^* \leq 360°$. This implies that for R = Hg and Xe, the most widespread configurations of collisions stabilizing CsBr are acute-angled triangular ones, i.e., energy is transferred to the third body, to some extent or another, in an interaction of R with *both* the ions. For none of the pairs $(E_i, E_R)$ with R = Hg or Xe, the collision configuration most favorable for energy transfer turns out to be collinear. For R = Kr, the deepest stabilization of the product molecules is attained, as a rule, in the same interval of the $\Theta$ angle. From the viewpoint of energy transfer, the most favorable collision configurations are those with $40° \leq \Theta \leq 70°$ (i.e., acute-angled triangular ones). Only for high energies $E_R$ of the third body, $8 \leq E_R \leq 10$ eV, the $\Theta$ angles between 0° and 30° become the most favorable ones: the deepest stabilization is attained in collisions of the Kr atom with the Br$^-$ ion in a collinear or nearly collinear configuration. To explain the presence of one peculiarity or another in the dependences $\Theta^*(E_i, E_R)$ and $\Phi^*(E_i, E_R)$, one has to examine the detailed dynamics of the process (cf. Refs. [13,14,22]) for each R.

## 4. Discussion

The differences in the behavior of Hg, Xe, and Kr as the third bodies in recombination of the atomic ions Cs$^+$ and Br$^-$ suggest that, most probably, the effectiveness of energy removal from the recombining pair depends not only on the mass of the third body but also on the interaction potentials of each of the atoms in question with the recombining particles. This conclusion is confirmed by the fact that the mean effectivities of the xenon and mercury atoms turn out to be almost the same despite a considerable difference in the masses. Almost equal effectivities of Xe and Hg are also seen while comparing the calculated minimal internal energies $E_m$ of the product molecules for most pairs $(E_i, E_R)$. This allows one to conjecture that the effectiveness of energy acceptance is determined not only by the masses of the Hg, Xe, and Kr atoms but also by the features of interaction dynamics of these atoms (as the third bodies) with the pair of the $Cs^+$ and $Br^-$ ions.

That both the factors are significant follows also from the qualitative and quantitative differences between the effectivity functions $E_m(E_i, E_R)$ for R = Hg, Xe (Figs. 1–2) and that for R = Kr (Fig. 3). For R = Hg (Fig. 1), the minimal internal energy $E_m$ of the products does not exceed 0.7 eV for any $E_R$ and is almost independent of the ion encounter energy $E_i$. Xenon atoms whose mass is 1.53 times smaller exhibit the same effectivity (Fig. 2). For krypton atoms (Fig. 3) which are 2.39 times lighter than Hg atoms and 1.57 times lighter than Xe atoms, the function $E_m(E_i, E_R)$ differs qualitatively from the corresponding functions for Hg and Xe. For most pairs $(E_i, E_R)$, CsBr molecules for R = Kr are stabilized to energies $E_m$ much higher than the minimal internal energies for R = Hg and Xe. It is also worthwhile to note that over the $E_R$ and $E_i$ ranges sampled, the deepest stabilization for R = Kr occurs, as a rule, at much smaller impact parameters $b_R = b_R^*$ than for R = Hg and Xe (see Figs. 4–6). One can therefore conclude that the differences in the behavior of the dependences $E_m(E_i, E_R)$ and $b_R^*(E_i, E_R)$ for R = Hg, Xe, Kr are due not only to the masses of the third bodies but also to the peculiarities of the interactions of these bodies with the recombining pair.

The approach employed in this paper is based on optimizing the kinematic parameters and does not enable one to separate *directly* the two factors determining the effectiveness of the third body in recombination processes (i.e., to "decouple" the mass effects from those of the characteristics of the interaction potentials of the third body with each of the recombining particles). However, this approach allows one to figure out how the effectivity function $E_m(E_i, E_R)$ and the mean effectivity $S$ change if one modifies formally the mass of the third body while the PES remains the same. Table 2 presents the values of all the nine mean effectivities obtained by considering the third bodies of the masses of the Hg, Xe, and Kr atoms in the framework of each of the three potential energy surfaces (see the end of Section 2). This table shows that for each of the PESs used, the averaged effectiveness attenuates strongly as the mass of the third body decreases. The PES most sensitive to the mass of the third body is the potential energy surface for the $Cs^+ + Br^- + Kr$ system. The data of Table 2 confirm again that the interactions of the third body with the individual ions as well as with the ion pair bound by the Coulomb potential play an important role. It is clear that such interactions are more intricate than the hard sphere model suggests. Of course, they are determined not only by the third body mass as one of the kinematic parameters but also by the topography of the PES that governs the elementary process in question.

**Table 2**
The mean effectivities (in eV) of the Hg, Xe, and Kr atoms calculated using the three potential energy surfaces governing the interactions $Cs^+ - Br^- - R$.

| The PES | Hg | Xe | Kr |
|---|---|---|---|
| $Cs^+ - Br^- - Hg$ | 0.16 | 0.21 | 0.46 |
| $Cs^+ - Br^- - Xe$ | 0.1 | 0.15 | 0.68 |
| $Cs^+ - Br^- - Kr$ | 0.07 | 0.16 | 0.71 |

Figs. 7 and 8 present the effectivity functions $E_m(E_i, E_R)$ for the PES of the $Cs^+ + Br^- + Kr$ system with "hypothetical" masses of the third body equal to the masses of the Xe and Hg atoms, respectively. The corresponding mean effectivities are $^{Xe}S_{Kr} = 0.16$ eV and $^{Hg}S_{Kr} = 0.07$ eV. A comparison of the effectivity functions shown in Figs. 7 and 8 with the function $E_m(E_i, E_R)$ of Fig. 3 proves that the structure of the effectivity function changes drastically and qualitatively as one replaces the Kr mass by the Xe mass and then by the Hg mass. As the mass of the third body increases from 83.8 to 131.29 Da, the maximal values of $E_m$ decrease from 2.6–2.7 to ≈1.6 eV, while the region where these maximal values are attained moves from $E_i \approx E_R \approx 9$–10 eV in Fig. 3 to $E_i \approx 1$ eV, $E_R \approx 10$ eV in Fig. 7. The topography of the effectivity function is altered significantly only for high energies $E_i$ and $E_R$. It is interesting to note that for low $E_i$ and high $E_R$, the structure of the effectivity function becomes similar to that of the function $E_m(E_i, E_R)$ of Fig. 2 calculated for the same mass of the third body but for the PES of the $Cs^+ + Br^- + Xe$ system. In the region of low energies $E_R$ from 1 to 4 eV, the effectivities $E_m$ remain almost the same as one passes from Fig. 3 to Fig. 7, the property of being independent of the ion encounter energy $E_i$ is also preserved to a considerable extent. A further increase in the third body mass to 200.59 Da (Fig. 8) leads to changes in the effectivity function that are at least as drastic as those just discussed. The averaged effectiveness of stabilization of the nascent molecule is further enhanced, and the maximal energy $E_m$ does not exceed 0.4 eV with the mean effectivity of 0.07 eV (the smallest value of all the nine mean effectivities $S$ given in Table 2). The general shape of the function $E_m(E_i, E_R)$ is very dissimilar to those presented in Figs. 3 and 7. The effectivity function in Fig. 8 exhibits only a not so pronounced maximum for high energies $E_i$ and $E_R$ and a sizeable region on the $(E_i, E_R)$ plane where the minimal internal energy of the CsBr molecules does not exceed 0.1 eV.

The situation with the PES for the $Cs^+ + Br^- + Hg$ system is different. As one replaces the third body of the Hg mass (see Fig. 1) by that of the Xe mass, the changes in the effectivity function $E_m(E_i, E_R)$ turn out to be rather minor. The most noticeable change is a slight increase in $E_m$ in the region of high energies $E_R$ and low energies $E_i$, so that the topography of the modified effectivity function resembles that of the function shown in Fig. 2. However, a further decrease in the mass of the third body to 83.8 Da drops the averaged effectiveness (see Table 2) and affects greatly the structure of the effectivity function where a well pronounced ravine-like minimum at $E_R \approx 5$ eV appears.

The changes in the effectivity function $E_m(E_i, E_R)$ for the PES of the $Cs^+ + Br^- + Xe$ system as one replaces the Xe mass of the third body by the Kr or Hg masses are analogous on the whole to those already discussed. The effectivity function calculated for this PES and the third body mass equal to 200.59 Da is presented in Fig. 9. This function is qualitatively similar more or less to that shown in Fig. 1.

It is worthwhile to emphasize that the structures of the effectivity functions for the collisions $Cs^+ + Br^- + Hg$ and $Cs^+ + Br^- + Xe$ (Figs. 1 and 2) are similar despite the fact that the PESs of these systems are constructed by rather unlike methods (see Section 2). The structure of the effectivity function for the collisions $Cs^+ + Br^- + Kr$ (Fig. 3) is different although the PESs for R = Xe and R = Kr are described by the same analytic expression (2) and differ only in the parameter values. On the other hand, as one decreases the third body mass from 131.29 to 83.8 Da without altering the PES, the changes in the effectivity function turn out to be drastic.

The qualitative differences in the effectiveness of R = Kr and of the two other atoms as the third bodies in the reactions (1) are probably due to the features of the interaction potentials of the Kr, Xe, and Hg atoms with the recombining ions $Cs^+$ and $Br^-$ bound by the Coulomb attraction. The potential energy curves $U(r)$ of the corresponding interactions are presented in Fig. 10.

Fig. 10 implies that strong repulsion between the ions and each of the atoms Hg and Xe occurs within approximately the same regions on the plane "the potential energy $U$ of repulsion – the internuclear distance $r$ between the interacting particles". For both the atoms, this region corresponds to a transfer of a considerable amount of energy from the recombining particles to the third body as the ion and the atom approach each other not so closely. Fig. 10 also shows that for the two heavy atoms R, the dependences of the repulsion energy $U$ on the distance $r$ "overlap". Most probably, it is these facts that ensure, to a great extent, the closeness of the effectivities of Hg and Xe as the third bodies. One may also note that at small distances $r$, the potential energy curves $U(r)$ in Fig. 10 for R = Hg and Xe are rather similar to potentials describing the hard sphere interaction.

On the other hand, the potential energy curves $U(r)$ for the $Cs^+$–Kr and $Br^-$–Kr repulsion presented in Fig. 10 suggest that in the case of R = Kr, the repulsion energies typical for the two heavy atoms are attained at much smaller distances $r$ of the approach of the krypton atom and the $Cs^+$ or $Br^-$ ions. Such distances $r$ are almost inaccessible for the collision energies sampled. This implies that at any given distance $r$ between the atom and the ion, energy transfer with R = Kr is less effective (on the average) for deep stabilization of the CsBr molecule than that with R = Xe or Hg. The differences in the energy transfer effectivities connected with the orientation effects could be yielded, to a great extent, by the peculiarities of the closest approach of the atom and the corresponding ions.

The present study does not consider the recombination reaction (1) for R = Ar due to the absence of suitable dynamical experiments and, consequently, of an adequate PES. However, a comparison of the interaction potentials for $Cs^+$–Ar, $Cs^+$–Kr, and $Cs^+$–Xe obtained from the mobilities of the $Cs^+$ ions measured in the Ar, Kr, and Xe gases at 300 K [34] shows that for any given internuclear distance between the particles, the $Cs^+$–Ar repulsion energy is much smaller than the $Cs^+$–Kr and $Cs^+$–Xe repulsion energies. One can therefore expect with great certainty that due to differences in the PES topography, the effectiveness of argon atoms as the third bodies is much lower than that of krypton and a fortiori xenon atoms.

Besides the mass of the third body and the interaction potential of the third body with each of the ions, there is one more factor affecting strongly the effectiveness of the third body as an energy acceptor in the triatomic systems in question. This factor of dynamical nature is the order of the interactions of the third body with the recombining pair. As was mentioned in the introduction, the energy transfer can happen via the interaction of the third body with both the ions simultaneously or with the two ions separately in any sequence. With the highest probability, the most effective stabilization of the nascent molecule occurs in the interaction of the third body with *both* the recombination partners. If this is the case, an acute-angled triangular configuration may be formed at the closest approach of the particles which is indeed typical for R = Hg, Xe and for most of the calculations with R = Kr. It is clear that one can observe a multitude of different triangular configurations which depend on the sequence of the interactions of the neutral atom with the ions. These configurations control the amount of energy transferred in every pairwise encounter. In turn, this amount of energy depends on the relation between the mass of the third body and the mass of the recombining particle the third body is interacting with. The presence of effective collinear collisions for R = Kr at high energies $E_R$ (see the end of Section 3) suggests that in the absence of acute-angled triangular configurations, only collinear or nearly collinear configurations can yield recombination with a sufficiently low internal energy of the molecule. The minimal internal energies of the CsBr molecules formed in this case are among the highest possible $E_m$ values.

Thus, the product molecule can be very effectively stabilized in simultaneous or consecutive encounters of the third body with both the recombining ions. It is worthwhile to note that such an effective stabilization does not necessarily require an approach of the recombining particles to the minimal distances. In principle, the nascent molecule can be stabilized at almost any stage of the existence of the bound pair of these particles [17]. For ions bound by the Coulomb interaction, the interionic distances at the stabilization instant could be controlled by the bond energy comparable with k$T$. To reveal more features of such processes, one has to examine detailed dynamics of recombination (cf. Refs. [13,14,22]).

## 5. Conclusion

The studies of the kinematic conditions for the elementary process of one-stage direct three-body recombination of the Cs$^+$ and Br$^-$ ions in the presence of the Hg, Xe, and Kr atoms have shown that the deepest stabilization of the products in the case of central encounters of the ions is attained for the mercury and xenon atoms. Their mean effectivities for the ion encounter energy $E_i$ ranging from 1 to 10 eV and the third body energy $E_R$ ranging from 1 to 10 eV are almost the same, and the corresponding effectivity functions $E_m(E_i, E_R)$ exhibit a rise in effectiveness on the whole as the third body energy increases, the dependence on the ion encounter energy being weak. The krypton atom effectiveness decreases as the third body energy grows.

The effectiveness of the third body R in the processes in question is determined mainly by three factors which cannot be separated sufficiently completely.

First, within the framework of each of the three PESs used for the Cs$^+$ + Br$^-$ + R systems, one observes an effect of the mass of the third body on the effectivity function and mean effectivity. This effect manifests itself most strongly for the PES of the Cs$^+$ + Br$^-$ + Kr system as one increases the third body mass to the mercury atom mass.

Another important factor is the structure of the potential energy surface and first of all the hardness of the potentials of the repulsive pairwise interactions between the third body and the recombining particles. This factor controls energy transfer at the distances (accessible for a given collision) of the closest approach of the particles. It is this factor that is the main reason for almost the same effectivities of the Hg and Xe atoms (the potential curves of the repulsion between these atoms and the recombining ions are similar).

The dynamical behavior of the three-body system during the collision gives rise to the third factor. The reaction mechanism determines the order of the interactions of the third body with the recombination partners, in particular, the number of encounters among the reagents and the particle configuration that yields recombination at given kinematic conditions of the collision. This dynamical factor probably includes the problem of the mass relation between the third body R and the recombination partner which encounter with R leads to transfer of energy sufficient to stabilize the product. Such a partner could be one of the recombining particles or both the particles simultaneously. In principle, this problem may be solved as one treats three-body recombination in the hard sphere approximation.

The approach employed in this study for examining three-body collisions can be used also for determining the kinematic conditions optimal for the formation of the recombination products in prescribed internal states. Moreover, one can apply the same approach to exploring the effectiveness of the third body in recombination of neutral particles provided that adequate PESs are available.

**Acknowledgment**

The authors are very grateful to M.B. Sevryuk for help in their research and fruitful discussions of the results obtained.

**Figure captions**

**Fig. 1.** The effectivity function $E_m(E_i, E_R)$ of the deep stabilization of the CsBr molecules for the direct three-body recombination $Cs^+ + Br^- + Hg \rightarrow CsBr + Hg$.

**Fig. 2.** The effectivity function $E_m(E_i, E_R)$ of the deep stabilization of the CsBr molecules for the direct three-body recombination $Cs^+ + Br^- + Xe \rightarrow CsBr + Xe$.

**Fig. 3.** The effectivity function $E_m(E_i, E_R)$ of the deep stabilization of the CsBr molecules for the direct three-body recombination $Cs^+ + Br^- + Kr \rightarrow CsBr + Kr$.

**Fig. 4.** The optimal impact parameters $b_R^*(E_i, E_R)$ in the direct three-body recombination $Cs^+ + Br^- + Hg \rightarrow CsBr + Hg$.

**Fig. 5.** The optimal impact parameters $b_R^*(E_i, E_R)$ in the direct three-body recombination $Cs^+ + Br^- + Xe \rightarrow CsBr + Xe$.

**Fig. 6.** The optimal impact parameters $b_R^*(E_i, E_R)$ in the direct three-body recombination $Cs^+ + Br^- + Kr \rightarrow CsBr + Kr$.

**Fig. 7.** The effectivity function for the PES of the $Cs^+ + Br^- + Kr$ system with the third body mass equal to the mass of the Xe atom.

**Fig. 8.** The effectivity function for the PES of the $Cs^+ + Br^- + Kr$ system with the third body mass equal to the mass of the Hg atom.

**Fig. 9.** The effectivity function for the PES of the $Cs^+ + Br^- + Xe$ system with the third body mass equal to the mass of the Hg atom.

**Fig. 10.** The pairwise interaction potentials of the $Cs^+$ and $Br^-$ ions with the Hg, Xe, and Kr atoms.

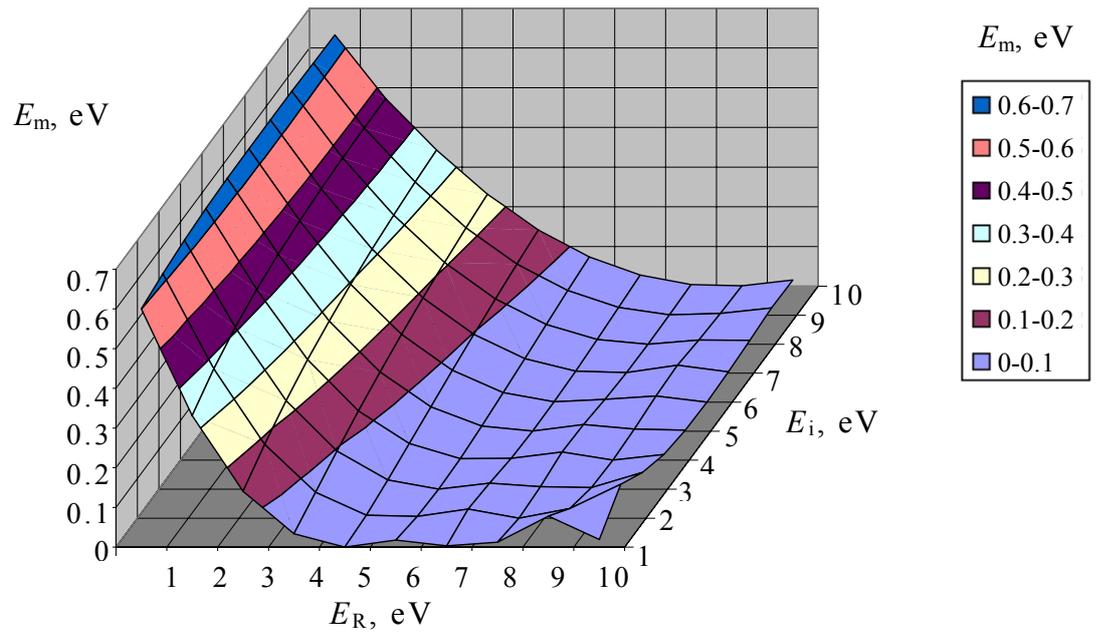

Fig. 1

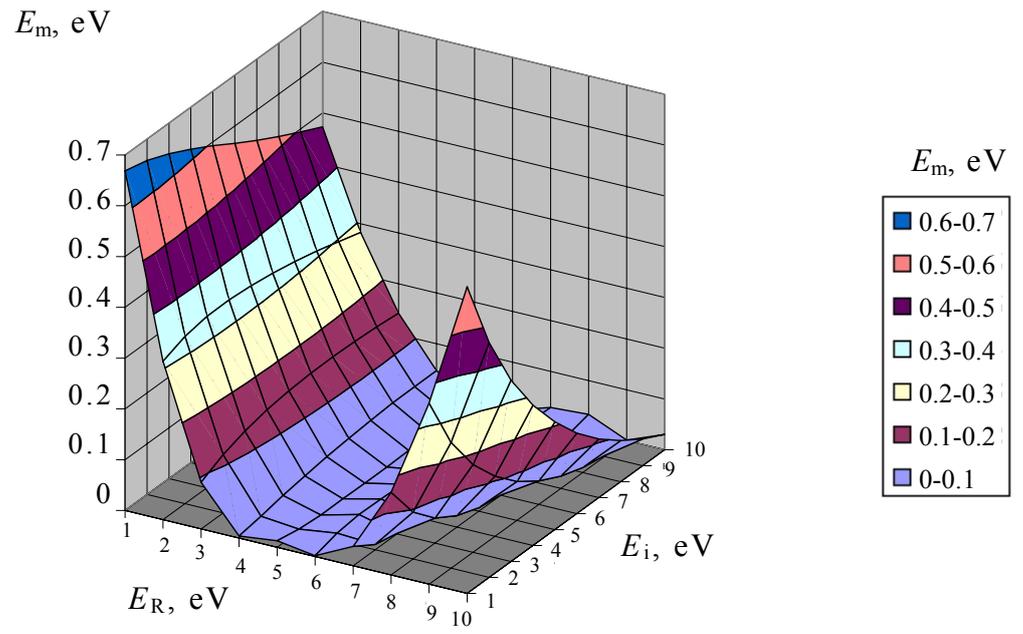

Fig. 2

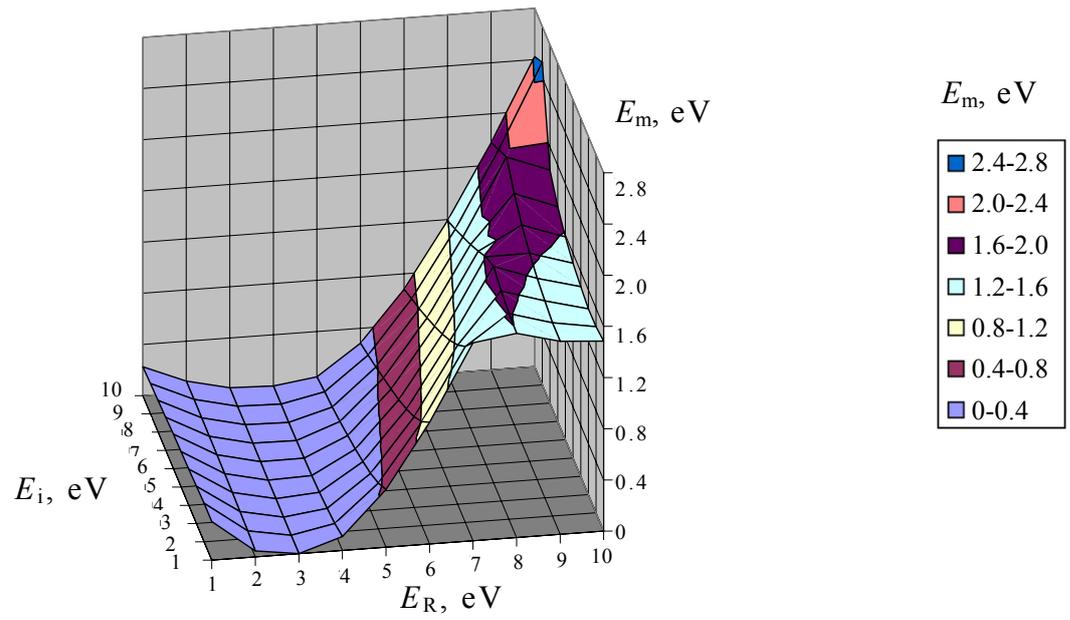

Fig. 3

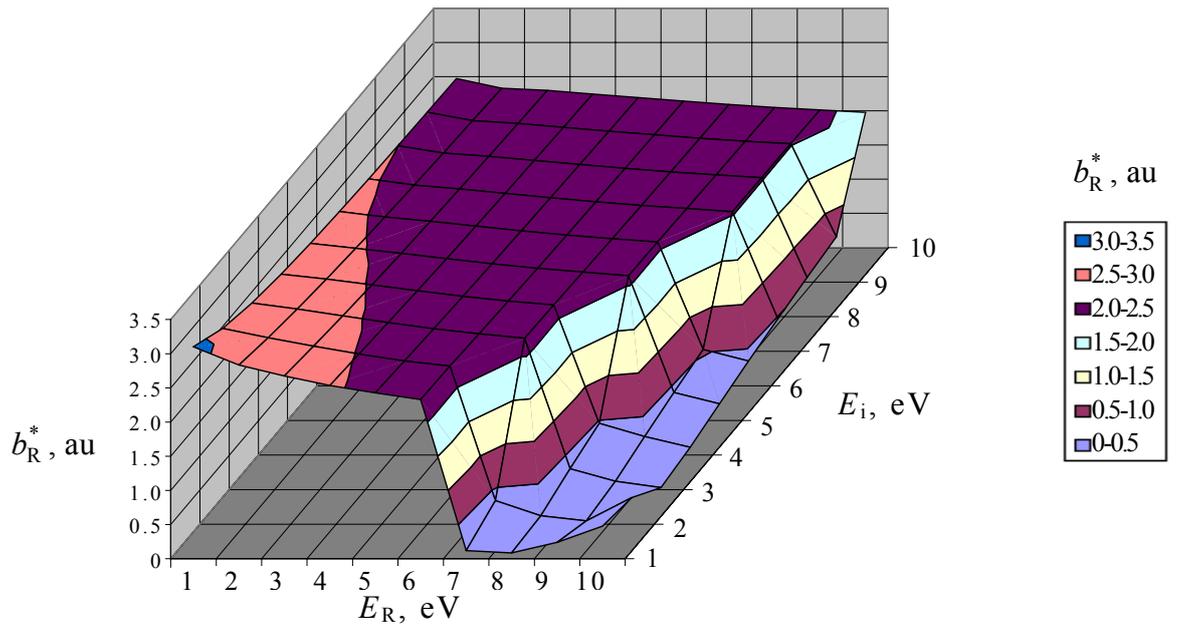

Fig. 4

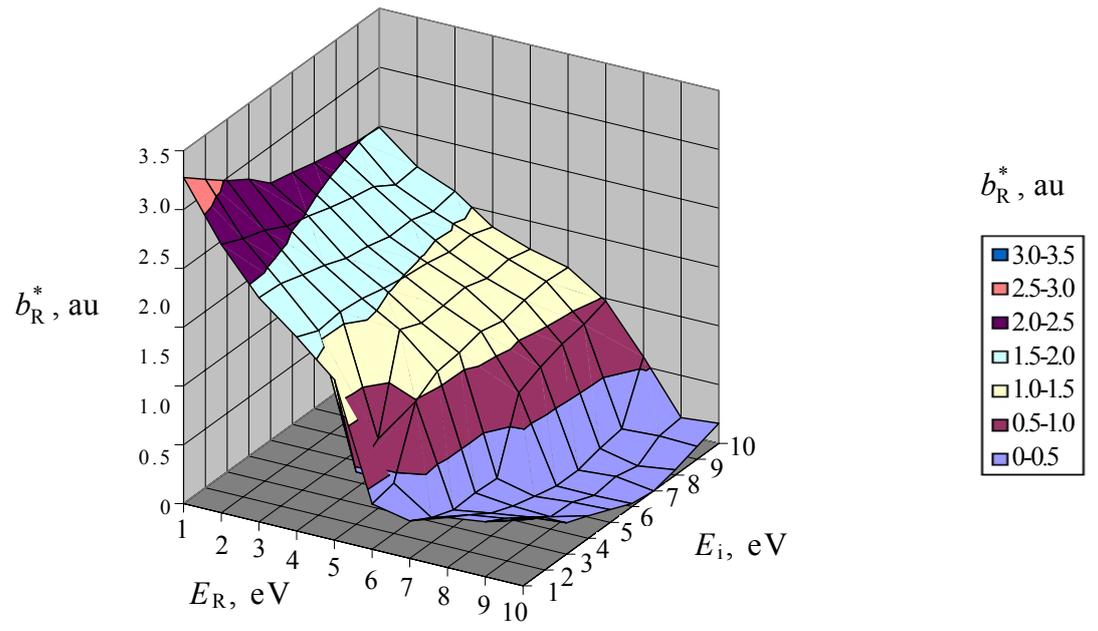

Fig. 5

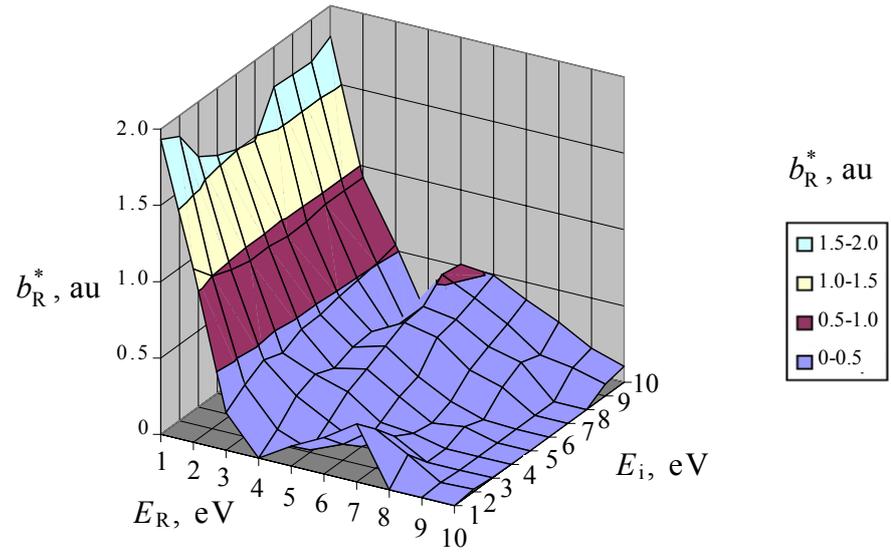

Fig. 6

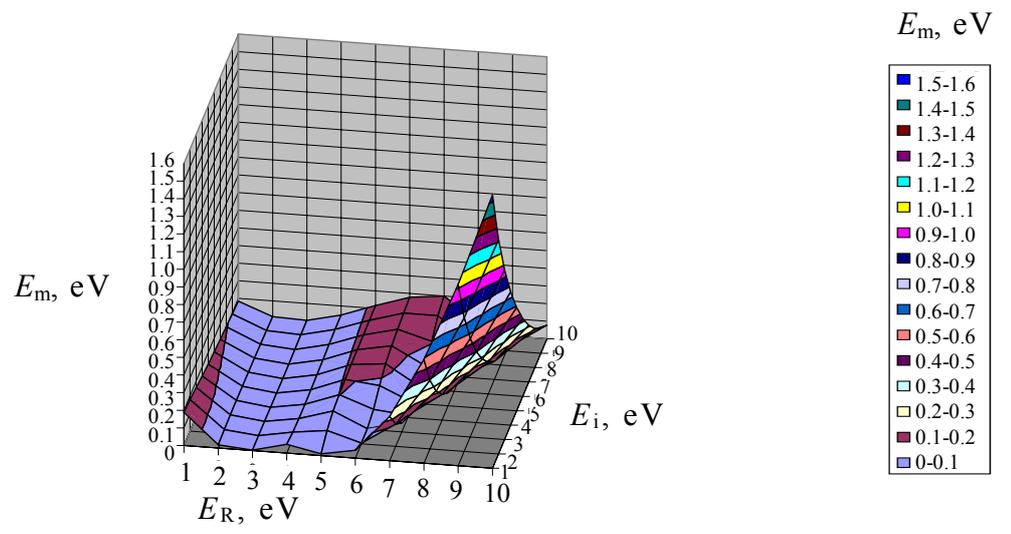

Fig. 7

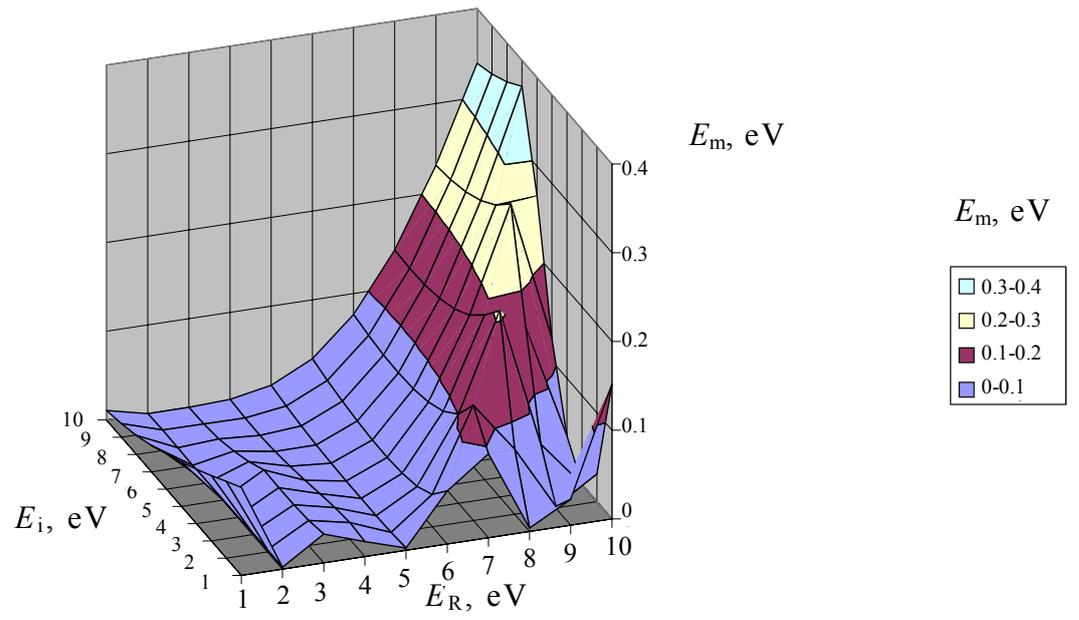

Fig. 8

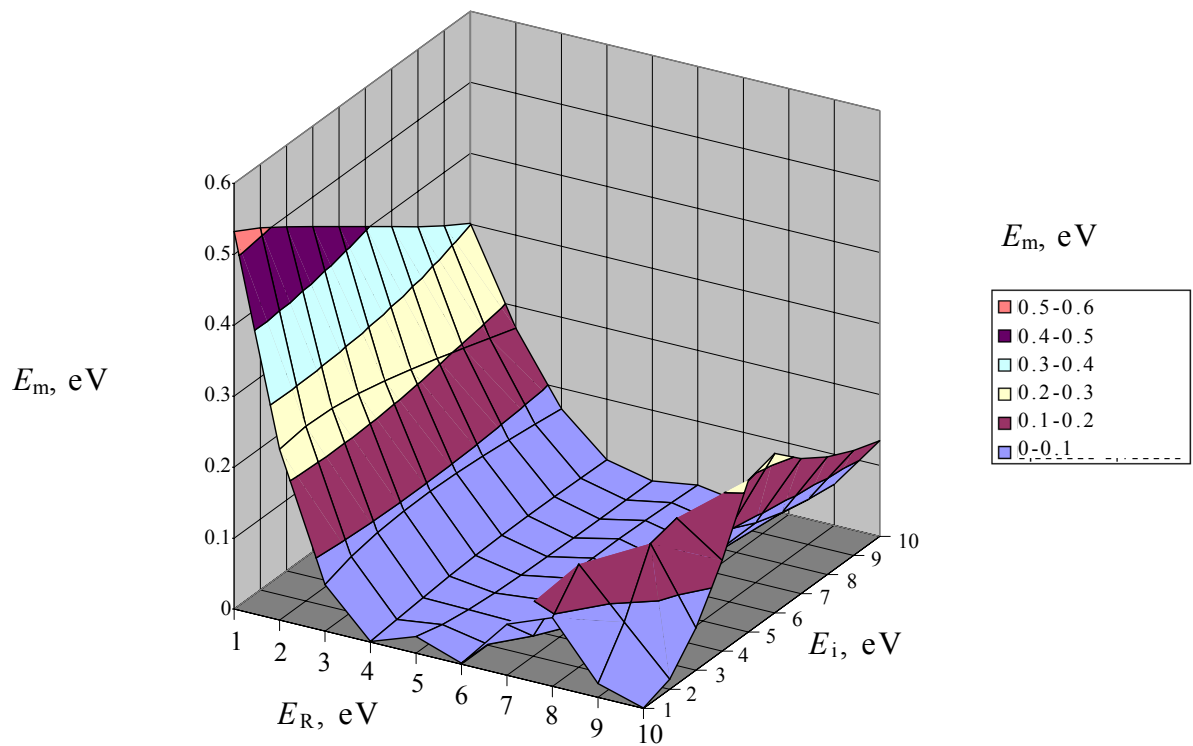

Fig. 9

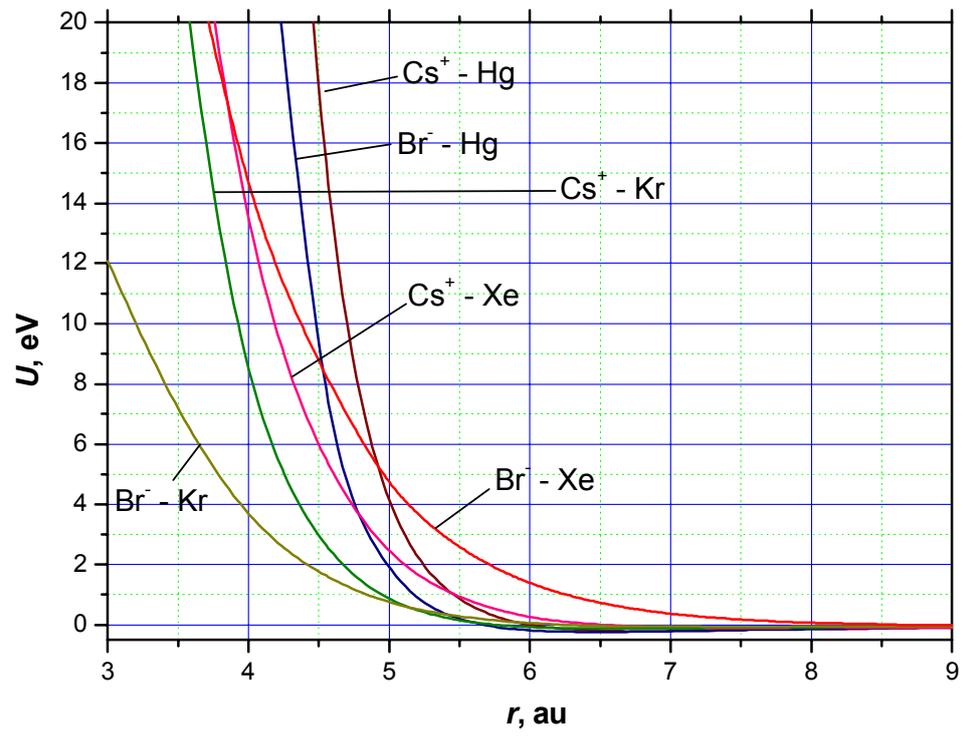

Fig. 10